\documentclass[aps,prd,twocolumn,showpacs,superscriptaddress,groupedaddress]{revtex4}  \usepackage{graphicx}  
\usepackage{dcolumn}   
\usepackage{bm}        
\usepackage{amssymb}   
\usepackage{amsmath}
\usepackage{array}
\usepackage{color}
\usepackage{overpic}

\newcommand{\PreserveBackslash}[1]{\let\temp=\\#1\let\\=\temp}
\newcolumntype{C}[1]{>{\PreserveBackslash\centering}p{#1}}
\newcolumntype{R}[1]{>{\PreserveBackslash\raggedleft}p{#1}}
\newcolumntype{L}[1]{>{\PreserveBackslash\raggedright}p{#1}}

\def \pipieta {\pi^{+}\pi^{-}\eta}

\begin{document}
\normalsize
\parskip=5pt plus 1pt minus 1pt

\title{\boldmath Evidence for $\eta_c$(2S)$\to\pipieta$ decay}

\author{
M.~Ablikim$^{1}$, M.~N.~Achasov$^{11,b}$, P.~Adlarson$^{70}$, M.~Albrecht$^{4}$, R.~Aliberti$^{31}$, A.~Amoroso$^{69A,69C}$, M.~R.~An$^{35}$, Q.~An$^{66,53}$, Y.~Bai$^{52}$, O.~Bakina$^{32}$
% ...
\\
\vspace{0.2cm}
(BESIII Collaboration)\\
\vspace{0.2cm} {\it
$^{1}$ Institute of High Energy Physics, Beijing 100049, People's Republic of China\\
% ...
$^{4}$ Bochum  Ruhr-University, D-44780 Bochum, Germany\\
% ...
$^{11}$ G.I. Budker Institute of Nuclear Physics SB RAS (BINP), Novosibirsk 630090, Russia\\
% ...
$^{31}$ Johannes Gutenberg University of Mainz, Johann-Joachim-Becher-Weg 45, D-55099 Mainz, Germany\\
$^{32}$ Joint Institute for Nuclear Research, 141980 Dubna, Moscow region, Russia\\
% ...
$^{35}$ Liaoning Normal University, Dalian 116029, People's Republic of China\\
% ...
$^{52}$ Southeast University, Nanjing 211100, People's Republic of China\\
$^{53}$ State Key Laboratory of Particle Detection and Electronics, Beijing 100049, Hefei 230026, People's Republic of China\\
% ...
$^{66}$ University of Science and Technology of China, Hefei 230026, People's Republic of China\\
% ...
$^{69}$ University of Turin and INFN, (A)University of Turin, I-10125, Turin, Italy; (B)University of Eastern Piedmont, I-15121, Alessandria, Italy; (C)INFN, I-10125, Turin, Italy\\
$^{70}$ Uppsala University, Box 516, SE-75120 Uppsala, Sweden\\
% ...
$^{b}$ Also at the Novosibirsk State University, Novosibirsk, 630090, Russia\\
% ...
}
}

\begin{abstract}
  The decay $\eta_c(2S)\to\pipieta$ is searched for through the radiative transition $\psi(3686) \to\gamma\eta_c(2S)$ using 448 million $\psi$(3686) events accumulated at the BESIII detector. The first evidence of $\eta_c(2S)\to\pi^+\pi^-\eta$ is found with a statistical significance of 3.5$\sigma$. The product of the branching fractions of $\psi(3686)\to\gamma\eta_c(2S)$ and $\eta_c(2S)\to\pipieta$ is measured to be $Br(\psi(3686)\to\gamma\eta_c(2S))\times Br(\eta_c(2S)\to\pipieta)=(2.97\pm0.81\pm0.26)\times10^{-6}$, where the first uncertainty is statistical and the second one is systematic. The branching fraction of the decay $\eta_c(2S)\to\pipieta$ is determined to be $Br(\eta_c(2S)\to\pipieta)=(42.4\pm11.6\pm3.8\pm30.3)\times10^{-4}$, where the third uncertainty is transferred from the uncertainty of the branching fraction of $\psi(3686)\to\gamma\eta_c(2S)$.

\end{abstract}
\maketitle

\section{Introduction}
Charmonium states play an important role in understanding the strong interaction, since their masses reside on the boundary between the perturbative and non-perturbative energy regions. The states below the open-charm threshold are better understood than the ones above, and their mass spectrum can be described by the quark potential model~\cite{Eichten:1978tg, Barnes:2005pb}. Our knowledge about the spin singlets, however, including the $P$-wave state $h_c$, the $S$-wave ground state $\eta_c$ and its first radial excitation $\eta_c(2S)$, is still limited~\cite{2011NBram}. The $\eta_c(2S)$ was first observed by the Belle collaboration via the decay $B^{\pm}\to K^{\pm}\eta_{c}(2S)$, $\eta_c(2S)\to K_{S}^{0}K^{\pm}\pi^{\mp}$ in 2002~\cite{2002Belle}, more than two decades after its prediction ~\cite{Buchmuller:1980su}. It was confirmed later by the CLEO~\cite{2004CLEO} and BaBar~\cite{2004BABAR} collaborations in the two-photon fusion process $\gamma\gamma\to\eta_c(2S)\to K^0_SK^{\pm}\pi^{\mp}$ and in addition by the BaBar collaboration in the double charmonium production process $e^+e^-\to J/\psi c\bar{c}$~\cite{2005BABAR}. The production of $\eta_c(2S)$ is also expected through the magnetic dipole (M1) transition of $\psi(3686)$~\cite{m1gamma}. The radiative transition $\psi$(3686)$\to\gamma\eta_c$(2S), with $\eta_c$(2S)$\to K^{+}K^{-}\pi^{0}$ and $K^{0}_{S}K^{\pm}\pi^{\mp}$, was reported by the BESIII collaboration in 2012~\cite{2012BESIII}.

The decay of charmonium states into light hadrons is believed to be dominated by the annihilation of the $c\bar{c}$ pair into two or three gluons. The so-called ``12\% rule'' states that the ratio of the inclusive branching fractions of light hadron final states between $\psi(3686)$ and $J/\psi$ is about 12\%~\cite{12percent}. Violations of this rule have been observed in various decay channels, especially in the $\psi\to\rho\pi$ process~\cite{Franklin:1983ve}. That so-called ``$\rho-\pi$ puzzle'' has not been solved yet. Similarly, one would expect a similar ratio of the hadronic branching fractions between $\eta_c(2S)$ and $\eta_c$ due to their analogous decaying dynamics in comparison to $\psi(3686)$ and $J/\psi$.
According to ref.~\cite{ansel}, for any normal light hadronic channel $h$,
\begin{equation}
\frac{Br(\eta_c(2S)\to h)}{Br(\eta_c\to h)}\approx\frac{Br(\psi(3686)\to h)}{Br(J/\psi\to h)} =0.128~,
\end{equation}
while ref.~\cite{chaokt} argues that this ratio should be close to one if no mixing with glueballs is considered. Recently, it was found that the experimental data significantly differ from both theoretical predictions~\cite{yuan}, \textit{i.e.}, most of the ratios are obviously greater than $12\%$ and less than one, except the ones with $p\bar{p}$ final states. Up to now, the total measured branching fraction of $\eta_c(2S)$ decays is small, \textit{i.e.}, less than 5\% according to the report from the particle data group (PDG), and the uncertainties of all the available experimental measurements are greater than 50\%~\cite{ParticleDataGroup:2020ssz}. Significantly improving the measurement precision of any single channel is difficult because of the still limited statistics of $\eta_c(2S)$ samples and, e.g., the difficulty in tagging the very soft radiative photon from the $\psi(3686)$ transition. Searching for more decay modes of the $\eta_c(2S)$ is therefore desirable to reduce the uncertainty of the averaged value of these ratios.

BESIII has collected (448.1$\pm$2.9)$\times$10$^6$ $\psi(3686)$ events in 2009 and 2012~\cite{sysnpsip}, large data samples that provide an excellent opportunity to search for new decay modes of the $\eta_c(2S)$. In this paper, the first evidence for the decay $\eta_c(2S)\to\pipieta$ via the M1 transition of $\psi(3686)$ is reported.

\section{BESIII EXPERIMENT and MONTE CARLO SIMULATION}
The BESIII detector~\cite{Ablikim:2009aa} records symmetric $e^+e^-$ collisions provided by the BEPCII storage ring~\cite{Yu:IPAC2016-TUYA01}, which operates in the center-of-mass energy range from 2.0 to 4.95~GeV, with a peak luminosity of $1 \times
10^{33}~\text{cm}^{-2}\text{s}^{-1}$ achieved at $\sqrt{s} = 3.77~\text{GeV}$.
BESIII has collected large data samples in this energy region~\cite{Ablikim:2019hff}. The cylindrical core of the BESIII detector covers 93\% of the full solid angle and consists of a helium-based  multilayer drift chamber~(MDC), a plastic scintillator time-of-flight system~(TOF), and a CsI(Tl) electromagnetic calorimeter~(EMC), which are all enclosed in a superconducting solenoid magnet providing a 1.0~T magnetic field. The solenoid is supported by an
octagonal flux-return yoke with resistive plate counter muon identification modules interleaved with steel. The charged-particle momentum resolution at $1~{\rm GeV}/c$ is $0.5\%$, and the d$E$/d$x$ resolution is $6\%$ for electrons from Bhabha scattering. The EMC measures photon energies with a resolution of $2.5\%$ ($5\%$) at $1$~GeV in the barrel (end cap) region. The time resolution in the TOF barrel region is 68~ps, while that in the end cap region is 110~ps.

Simulation samples produced with a {\sc geant4}-based~\cite{geant4} Monte Carlo (MC) package, which includes the geometric description of the BESIII detector and the detector response, are utilized to determine reconstruction efficiencies and to estimate background contributions. The simulation models the beam-energy spread and initial state radiation (ISR) in the $e^+e^-$ annihilation with the generator {\sc
kkmc}~\cite{ref:kkmc}. The inclusive MC sample includes the production of the $\psi(3686)$ resonance, the ISR production of the $J/\psi$, and the continuum processes incorporated in {\sc kkmc}~\cite{ref:kkmc}. The decay modes are modelled with {\sc evtgen}~\cite{ref:evtgen} using the known averaged branching fractions~~\cite{ParticleDataGroup:2020ssz}, and the unknown charmonium decays are modelled with {\sc lundcharm}~\cite{ref:lundcharm}. Final state radiation~(FSR) from charged final state particles is incorporated using the {\sc photos} package~\cite{photos}. Event type analysis of the inclusive MC samples with a generic tool, TopoAna~\cite{Topo}, is used to study the potential background. The exclusive decays of $\psi(3686)\to\gamma\rm{X}(\rm{X}=\chi_{c1,2}$ and $\eta_c(2S))$ are generated taking into the angular distribution, while the decay $\rm{X}\to\pipieta$ is generated uniformly in phase space (PHSP).

\section{Event selection and background analysis}
The $\eta_c(2S)$ candidates studied in this analysis are produced via the M1 transition of $\psi(3686)\to\gamma\eta_c(2S)$($\eta_c(2S)\to\pipieta$), the $\eta$ is reconstructed via its two-photon decay ($\eta\to\gamma\gamma$). Therefore, the final state is $\gamma\gamma\gamma\pi^+\pi^-$, {\it i.e.}, the candidate events are required to have two charged tracks with a net charge of zero, and at least three photons.

Charged tracks detected in the MDC are required to be within a polar angle ($\theta$) range of $|\rm{cos\theta}|<0.93$, where $\theta$ is defined with respect to the $z$-axis
which is the symmetry axis of the MDC. For charged tracks, the distance of the closest approach to the interaction point must be less than $10$~cm along the $z$-axis, and less than $1$~cm
in the transverse plane. Particle identification~(PID) for charged tracks combines measurements of the energy deposited in the MDC~(d$E$/d$x$) and the flight time in the TOF to form likelihoods $\mathcal{L}(h)~(h=p,K,\pi)$ for each hadron $h$ hypothesis. Tracks are identified as \mbox{pions} when the pion hypothesis has the greatest likelihood ($\mathcal{L}(\pi)>\mathcal{L}(K)$ and $\mathcal{L}(\pi)>\mathcal{L}(p)$). Each event is required to have one $\pi^+$ and one $\pi^-$.

Photon candidates are identified using showers in the EMC.  The deposited energy of each shower must be more than 25~MeV in the barrel region ($|\cos \theta|< 0.80$) and more than 40~MeV in the end cap region ($0.86 <|\cos \theta|< 0.92$). To exclude showers that originate from charged tracks, the angle subtended by the EMC shower and the position of the closest charged track at the EMC must be greater than 10 degrees as measured from the interaction point. To suppress electronic noise and showers unrelated to the event, the difference between the EMC time and the event start time is required to be within [0, 700] ns.

A kinematic fit with five constraints (5C) on each candidate is performed, where the total energy-momentum of final states is constrained to the initial four-momentum and the invariant mass of the two photons is constrained to the nominal $\eta$ mass. The fit loops over all possible $\eta\to\gamma\gamma$ and M1 $\gamma$ combinations, and the combination with the minimal $\chi^2_{5C}(3\gamma)$ is selected. In order to suppress the background channels $\psi$(3686)$\to\pipieta$ and $\psi$(3686)$\to\gamma\gamma\pipieta$,  with one less or one more photon in the final state than the signal, the value of $\chi^2_{5C}(3\gamma)$ is required to be less than $\chi^2_{5C}(2\gamma)$ and $\chi^2_{5C}(4\gamma)$. Also, a selection is set on the $\chi^2_{5C}(3\gamma)$ distribution.

\begin{figure}[htbp]
\begin{center}
\begin{overpic}[width=0.47\textwidth]{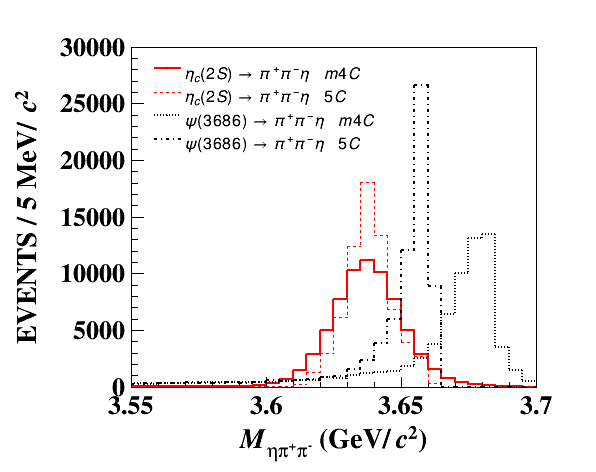}
\end{overpic}
\caption{The comparison of $M_{\pipieta}$ obtained from the m4C and 5C kinematic fits of MC samples (Color Online). The red solid curve and the red dashed curve denote the m4C and 5C kinematic results of the signal channel, respectively; the black dotted curve and the black dash-dotted curve denote the m4C and 5C kinematic results of $\psi$(3686)$\to\pipieta$, respectively.}
\label{fig::comp_mpipieta}
\end{center}
\end{figure}

 As can be seen in the invariant mass spectrum of $\pipieta$  in Fig.~\ref{fig::comp_mpipieta}, the background contribution of the channel $\psi(3686)\to\pipieta$ with a fake photon would appear as a peak close to the $\eta_c(2S)$ signal. This peak can be reduced, if the energy of the candidate M1 photon is not used in the kinematic fit, by adjusting the 5C fit to a modified 4C fit (m4C)~\cite{2012BESIII}. By this method, this background channel is reduced in the $\eta_c(2S)$ mass region as shown in Fig.~\ref{fig::comp_mpipieta}. In this analysis, the invariant mass distribution after applying the m4C fit is used for further study.

\begin{figure}[htbp]
\begin{center}
\begin{overpic}[width=0.47\textwidth]{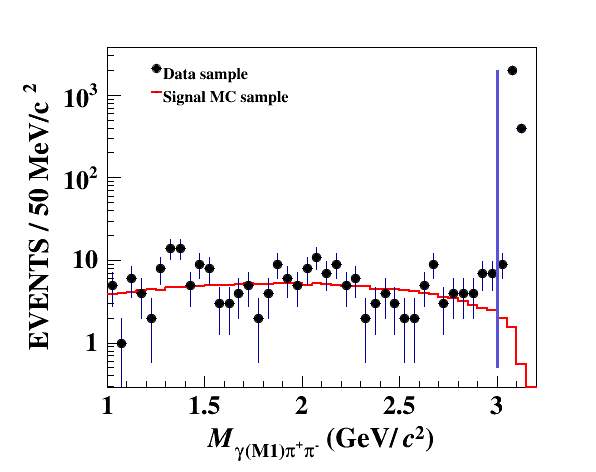}
\end{overpic}
\caption{The distributions of $M_{\gamma_{\text{M1}}\pi^+\pi^-}$ of data and signal MC samples. The black dots with error bars denote the data sample, the red histogram denotes the signal MC sample, and the blue vertical line indicates the corresponding selection criterion.}
\label{fig::comp_mgpipi}
\end{center}
\end{figure}

The signal suffers significantly from background contributions associated with $J/\psi$ decaying to $\mu^+\mu^-(\gamma_{\rm FSR})$, such as $\psi(3686)\to\eta J/\psi$ and $\pi^0\pi^0 J/\psi$. As shown in Fig.~\ref{fig::comp_mgpipi}, there is a large enhancement around the $J/\psi$ resonance in the invariant mass of ${\gamma(\text{M1})\pi^+\pi^-}$. Events with $M_{\gamma(\text{M1})\pi^+\pi^-}\leq$ 3.00 GeV/$c^2$ are accepted for further study.

The decay of $\psi(3686)\to\pipieta\gamma_{\rm FSR}$ with an FSR photon cannot be reduced because of the same final state as the signal decay mode. It could potentially contaminate the signal channel with a long tail in the invariant mass of ${\pipieta}$. The contribution of this channel depends on the FSR ratio ($R_{\rm FSR}$), which is defined as $R_{\rm FSR}=\frac{N_{\rm FSR}}{N_{\rm no\,FSR}}$ ignoring any kinematic dependence, where $N_{\rm FSR}$ and $N_{\rm no\,FSR}$ are the numbers of events with and without an FSR photon~\cite{fsrest}. A control sample of $\psi(3686)\to\gamma\chi_{c0}\to\gamma\gamma_{\rm FSR}2(\pi^+\pi^-)$ is selected to study the difference of $R_{\rm FSR}$ between data and MC sample. From the analysis, we find $f_{\rm FSR}=R_{\rm FSR}^{\rm Data}/R_{\rm FSR}^{\rm MC}=1.70\pm0.07$, with a purely statistical uncertainty. The line-shape of $\psi(3686)\to(\gamma_{\rm FSR})\pipieta$ is described by the sum of MC simulated shapes of $\psi(3686)\to\pipieta$ and $\psi(3686)\to\gamma_{\rm FSR}\pipieta$ with the FSR ratio corrected by the factor $f_{\rm FSR}$. It is then used to describe the corresponding background contribution later in the fit.

In addition, background channels such as $\psi(3686)\to\gamma\chi_{c1}(\pi^+\pi^-\pi^0)$ and $\psi$(3686)$\to\omega(\gamma\pi^{0})\pi^{+}\pi^{-}$ with a $\pi^0$ in the final state contribute only insignificantly. Furthermore, the line shapes of these two decay modes are smooth in the invariant mass of ${\pipieta}$ according to the study of MC samples, so no $\pi^0$ veto is applied to ensure the significance of the signal.

\section{Branching fraction determination}
The branching fraction is calculated by

\begin{equation}
Br(\eta_c(2S)\to\pipieta)=\frac{N_{\rm sig}}{N_{\psi(3686)}\times\epsilon\times BR_1\times BR_2},
\end{equation}
where $N_{\rm sig}$ is the observed number of signal events, $N_{\psi(3686)}$ is the total number of $\psi(3686)$ decays~\cite{sysnpsip}, $\epsilon$ is the detection efficiency, $BR_1$ and $BR_2$ are the branching fractions of $\psi(3686)\to\gamma\eta_c(2S)$ and $\eta\to\gamma\gamma$~\cite{ParticleDataGroup:2020ssz}, respectively. These values are listed in Table~\ref{tab:input}.

\begin{table}[htbp]
 \centering
 \caption{Input values for calculating the branching fraction $Br(\eta_c(2S)\to\pipieta)$. \label{tab:input}}
\begin{tabular}{ccccc}
\hline\hline
$N_{sig}$ & $N_{\psi(3686)}(\times10^6)$ & $\epsilon$ & $BR_1(\times10^{-4})$ & $BR_2(\times10^{-2})$  \\ \hline
$106\pm29$ & $448.1\pm2.9$ & 0.202 & $7\pm5$ & $39.41\pm0.20$ \\
\hline\hline
\end{tabular}
\end{table}

\begin{figure}[htbp]
\begin{center}
\begin{overpic}[width=0.47\textwidth]{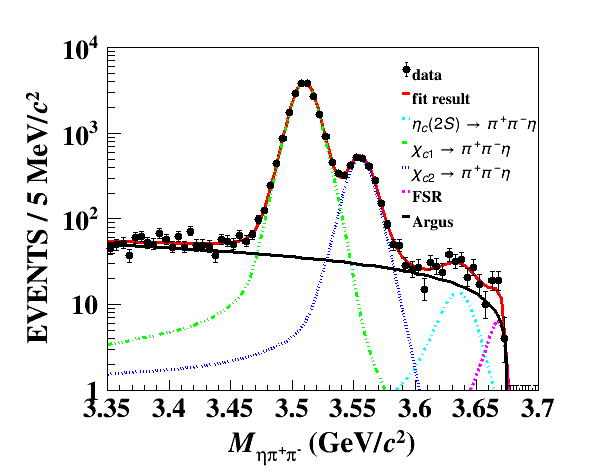}
\end{overpic}
\caption{The result of a fit to the invariant mass distribution of ${\pi^+ \pi^- \eta}$. The black dots with error bars are BESIII data, the red and black solid curves denote the total fit curve and the shape of the smooth background contributions, respectively. The green dash-dot-dotted curve, the blue dotted, and the cyan dash-dotted denote the decay modes of $\chi_{c1}$, $\chi_{c2}$ and $\eta_c(2S)$, while the pink dashed curve denotes the contributions of the FSR process $\psi(3686)\to(\gamma_{\rm FSR})\pipieta$.}
\label{fig::fit_mean}
\end{center}
\end{figure}

The signal yield $N_{\rm sig}$ is extracted by an unbinned maximum likelihood fit to the invariant mass distribution of ${\pipieta}$ (see Fig.~\ref{fig::fit_mean}). The fit range is from 3.35 to 3.70 GeV/$c^2$, which includes the $\chi_{c1}$ and $\chi_{c2}$ signals. The line shapes of $\eta_c(2S)$ and $\chi_{c1,2}$ are described by the following formulas,
\begin{equation}
[E^{3}_{\gamma}\times BW_0(m)\times f_{d}(E_{\gamma})]\otimes DGaus(\delta m,\sigma) \,
\end{equation}
and
\begin{equation}
[E^{3}_{\gamma}\times (BW_1(m)+BW_2(m))]\otimes DGaus(\delta m,\sigma) \,
\end{equation}
respectively, Here, $m$ is the invariant mass of ${\pipieta}$, $E_{\gamma}=\frac{m^2_{\psi(3686)}-m^2}{2m_{\psi(3686)}}$ is the energy of the transition photon in the rest frame of $\psi(3686)$, $BW_0$, $BW_1$ and $BW_2$ denote the Breit-Wigner functions for $\eta_c(2S)$, $\chi_{c1}$ and $\chi_{c2}$ with the mass and width fixed to the reported averaged values of these three resonances~\cite{ParticleDataGroup:2020ssz}, $DGaus(\delta m,\sigma)$ is a double-Gaussian function describing the mass shift and resolution whose parameters are shared by $\chi_{cJ}$ and $\eta_c(2S)$, $f_d(E_{\gamma})$ is a damping function to suppress the diverging caused by  $E^3_\gamma$~\cite{kedrdumpf}, i.e., $f_d(E_{\gamma})=\frac{E^2_0}{E_{\gamma}E_0+(E_\gamma-E_0)^2}$ with $E_0=\frac{m^2_{\psi(3686)}-m^2_{\eta_c(2S)}}{2m_{\psi(3686)}}$ the mean energy of the transition photon chosen for this work. The contributions of $\chi_{c1,2}$ backgrounds are determined by the fit, and the results are consistent with the world averaged values~\cite{ParticleDataGroup:2020ssz}. The remaining background due to the FSR process $\psi(3686)\to(\gamma_{\rm FSR})\pipieta$ is described by the MC shape. All smooth background contributions are described by an Argus function with the threshold fixed and the other parameters floated.

The fit result is shown in Fig.~\ref{fig::fit_mean} and the goodness of fit is $\chi^2/ndf =67/59=1.1$, where $ndf$ denotes the number of degrees of freedom. The yield of the $\eta_c(2S)$ signal is $N_{\rm sig}=106\pm29$ with a statistical significance of 3.5$\sigma$ that is obtained from the difference of the logarithmic likelihoods~\cite{signf}, taking into account the difference of $ndf$. We obtain a product branching fraction of $Br(\psi(3686)\to\gamma\eta_c(2S))\times Br(\eta_c(2S)\to\pipieta)=(2.97\pm0.81)\times10^{-6}$, and a branching fraction of $Br(\eta_c(2S)\to\pipieta)=(42.4\pm11.6)\times10^{-4}$, in which only the statistical uncertainties are presented.

\section{Systematic uncertainties}
\label{sec:sys}
The sources of systematic uncertainties for the branching fraction measurement include the number of $\psi(3686)$ events, tracking, PID, photon detection, the kinematic fit, the branching fractions of the intermediate decays, and the fit to the invariant mass distribution of ${\pipieta}$.
\begin{itemize}
\item Number of $\psi(3686)$. The uncertainty due to the number of $\psi(3686)$ events is determined with inclusive hadronic $\psi(3686)$ decays and estimated to be 0.6\%~\cite{sysnpsip}.

\item Tracking and PID efficiency. The pion tracking efficiency was studied by using a control sample of $J/\psi\to p \bar{p} \pi^+\pi^-$. The difference in the tracking efficiencies between MC simulation and data is 1\% per pion~\cite{systrack}. The pion PID  efficiency was investigated with a control sample of $J/\psi\to\rho\pi$~\cite{syspid}. The difference in the PID efficiencies between data and MC simulation is 1\% per pion.

\item Photon detection. The uncertainty due to photon reconstruction is determined to be 1\% per photon by the study of the process $J/\psi\to\pi^+\pi^-\pi^0$~\cite{syspho}.

\item Kinematic fit. A control sample of $\psi(3686)\to\gamma\chi_{c1}$, $\chi_{c1}\to\pipieta$ is selected to estimate the uncertainty associated with the kinematic fit. The difference of the efficiencies with and without the kinematic fit is taken as the corresponding uncertainty, determined to be 1.6\%.

\item Branching fractions. The systematic uncertainties due to the branching fractions of $\psi(3686)\to\gamma\eta_c(2S)$ and $\eta\to\gamma\gamma$ are 71.4\% and 0.5\% according to the PDG~\cite{ParticleDataGroup:2020ssz}.

\item Fit to the invariant mass distribution of ${\pi^+\pi^-\eta}$. Seven potential sources of systematic uncertainties are considered. First, the uncertainty from the background shape is estimated by using an alternative Chebychev function. The change of the fitted signal yield, 3.8\%, is assigned as the uncertainty. Second, the uncertainty of the signal shape is estimated by changing the width of $\eta_c(2S)$ with $\pm$1 standard deviation away from the nominal value, and is determined to be 4.7\%. Third, the uncertainty caused by the shapes of $\chi_{c1}$ and $\chi_{c2}$ is estimated by changing the widths of $\chi_{c1}$ and $\chi_{c2}$ with $\pm$1 standard deviation away from their nominal values, it is determined to be 1.9\%. Fourth, the damping function is changed to an alternative form used by CLEO~\cite{cleodumpf}, $f_d(E_{\gamma})=$exp$(-\frac{E^2_{\gamma}}{8\beta^2})$ with $\beta=(65.0\pm2.5)$ MeV, the resulting difference in the fit is 3.8\%, and is assigned as the systematic uncertainty. Fifth, the systematic uncertainty from the double-Gaussian function is estimated by changing it to a triple-Gaussian function, and the resulting difference is 0.9\%. Sixth, the systematic uncertainty associated with the $f_{\rm FSR}$ is estimated by varying the $f_{\rm FSR}$ with $\pm$1 standard deviation that is mainly caused by the statistics of the control samples, and the difference is 0.9\%. Seventh, the fit range is varied, and the maximum differences in the fitted yields are considered as the associated systematic uncertainties. It is determined to be 1.9\%.

\end{itemize}
Among all sources of systematic uncertainties, by far the largest one comes from the branching fraction of $\psi(3686)\to\gamma\eta_c(2S)$, and is therefore treated separately. All other sources of systematic uncertainties are assumed to be independent of each other and combined in quadrature to obtain the overall systematic uncertainty as listed in Table~\ref{tab:sys}.
\begin{table}[htbp]
 \centering
 \caption{Summary of systematic uncertainties. \label{tab:sys}}
\begin{tabular}{lc}
\hline\hline
Source  & Uncertainty (\%)  \\ \hline
$N_{\psi(3686)}$             &0.6 \\
Tracking                   &2.0  \\
 PID                       &2.0  \\
 Photon                    &3.0 \\
 Kinematic fit             &1.6  \\
 $Br(\eta\to\gamma\gamma)$    &0.5   \\
 Background shape   &3.8 \\
Width of $\eta_c(2S)$  &4.7 \\
Width of $\chi_{c1,2}$ & 1.9\\
 Damping function   &3.8   \\
 Double-Gaussian function & 0.9 \\
 Ratio of FSR   &0.9 \\
 Fit range  &1.9 \\ \hline

 Total    & 8.9 \\ \hline
 $Br(\psi$(3686)$\to\gamma\eta_c$(2S))& 71.4  \\
\hline\hline
\end{tabular}
\end{table}
\section{Results and discussion}
With the (448.1$\pm$2.9)$\times$10$^6$ $\psi(3686)$ data sample, the process of $\eta_c(2S)\to\pipieta$ is searched for by utilizing the M1 transition $\psi(3686)\to\gamma \eta_c(2S)$. Evidence for the decay $\eta_c(2S)\to\pipieta$ is found for the first time, with a statistical significance of 3.5$\sigma$. The product of the branching fractions is measured to be $Br(\psi(3686)\to\gamma\eta_c(2S))\times Br(\eta_c(2S)\to\pipieta)=$  (2.97$\pm$0.81$\pm$0.26)$\times10^{-6}$, where the first uncertainty is statistical and the second systematic. The branching fraction of $\eta_c(2S)$ decaying into $\pi^+ \pi^- \eta$ is measured to be $Br(\eta_c(2S)\to\pipieta)=$  $(42.4\pm11.6\pm3.8\pm30.3)\times10^{-4}$, with the first uncertainty being the statistical, the second the systematic uncertainty without taking into account the uncertainty of the branching fraction of $\psi(3686)\to\gamma\eta_c(2S)$. The third one is the systematic uncertainty arising from this branching fraction.

\begin{figure}[htbp]
\begin{center}
\begin{overpic}[width=0.47\textwidth]{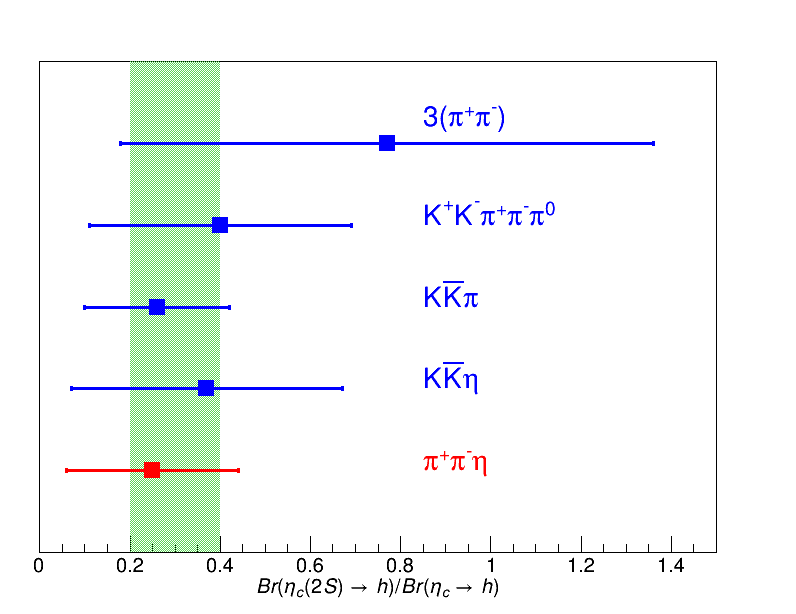}
\end{overpic}
\caption{Estimation of the averaged value of the ratio of Br($\eta_c(2S)\to h$) to Br($\eta_c\to h$). Here, $h$ means various hadronic final states, as shown on this figure. Except the branching fraction of $\eta_c(2S) \to \pi^+ \pi^- \eta$ of this work, the other results are quoted from Refs.~\cite{ParticleDataGroup:2020ssz,etacptospi}.  The shade is the averaged value of these five decay modes with one standard deviation. Both statistical and systematic uncertainties have been included.}
\label{fig::comp_br}
\end{center}
\end{figure}

With the branching fraction  $Br(\eta_c\to\pi^+\pi^-\eta)=$(1.7$\pm$0.5)\%~\cite{ParticleDataGroup:2020ssz}, the ratio of the branching fractions of $\eta_c$ and $\eta_c(2S)$ decaying into $\pi^+ \pi^- \eta$ is calculated to be $\frac{Br(\eta_c(2S)\to\pi^+\pi^-\eta)}{Br(\eta_c\to\pi^+\pi^-\eta)}=0.25\pm0.20$. Combining the ratios of other hadronic decay modes of $\eta_c(2S)$ to $\eta_c$~\cite{ParticleDataGroup:2020ssz,etacptospi}, the averaged value of all these ratios including this measurement is determined to be 0.30$\pm$0.10 (see Fig.~\ref{fig::comp_br}). This ratio agrees neither with the prediction in Ref.~\cite{ansel} nor in Ref.~\cite{chaokt}. The observed discrepancy reflects our limited knowledge of the decay mechanisms of the spin singlet charmonium states. More searches on new decay modes and more precise measurements of the $\eta_c(2S)$ decays are required to shed light on this puzzle. With about 3 billion $\psi(3686)$ events to be accumulated, BESIII will make a further substantial contribution to this field~\cite{besiiibook}. A better precision of the branching fraction of $\psi(3686) \to \gamma \eta_c(2S)$ is hereby crucial.

\acknowledgments
The BESIII collaboration thanks the staff of BEPCII and the IHEP computing center for their strong support. This work is supported in part by National Key R\&D Program of China under Contracts Nos. 2020YFA0406300, 2020YFA0406400; National Natural Science Foundation of China (NSFC) under Contracts Nos. 12275058, 11875115, 11635010, 11735014, 11835012, 11935015, 11935016, 11935018, 11961141012, 12022510, 12025502, 12035009, 12035013, 12192260, 12192261, 12192262, 12192263, 12192264, 12192265; the Chinese Academy of Sciences (CAS) Large-Scale Scientific Facility Program; Joint Large-Scale Scientific Facility Funds of the NSFC and CAS under Contract Nos. U1832207, U2032110; the CAS Center for Excellence in Particle Physics
(CCEPP); 100 Talents Program of CAS; The Institute of Nuclear and Particle Physics (INPAC) and Shanghai Key Laboratory for Particle Physics and Cosmology; ERC under Contract No. 758462; European Union's Horizon 2020 research and innovation programme under Marie Sklodowska-Curie grant agreement under Contract No. 894790; German Research Foundation DFG under Contracts Nos. 443159800, Collaborative Research Center CRC 1044, GRK 2149; Istituto Nazionale di Fisica Nucleare, Italy; Ministry of Development of Turkey under Contract No. DPT2006K-120470; National Science and Technology fund; National Science Research and Innovation Fund (NSRF) via the Program Management Unit for Human Resources \& Institutional Development, Research and Innovation under Contract No. B16F640076; Olle Engkvist Foundation under Contract No. 200-0605; STFC (United Kingdom); Suranaree University of Technology (SUT), Thailand Science Research and Innovation (TSRI), and National Science Research and Innovation Fund (NSRF) under Contract No. 160355; The Royal Society, UK under Contracts Nos. DH140054, DH160214; The Swedish Research Council; U. S. Department of Energy under Contract No. DE-FG02-05ER41374.

\end{document}